\documentclass[prd,aps,showpacs,twocolumn,amssymb,amsmath,amstext,amsfont,noshowkeys,groupedaddress]{revtex4}
\usepackage{graphicx}
\usepackage{theorem}
\usepackage{dcolumn}
\usepackage{bm}
\usepackage{mathrsfs}
\usepackage{color}

\begin{document}

\title{A Concept for A Dark Matter Detector Using Liquid Helium-4}
\author{W. Guo}
\email[Electronic address: ]{wguo@magnet.fsu.edu}
\affiliation{Mechanical engineering department, Florida State
University, Tallahassee, FL 32310, USA}
\author{D.N. McKinsey}
\email[Electronic address: ]{daniel.mckinsey@yale.edu}
\affiliation{Physics department, Yale University, New Haven, CT 06520, USA}
\date{\today}

\begin{abstract}
Direct searches for light dark matter particles (mass $<10$ GeV) are
especially challenging because of the low energies transferred in
elastic scattering to typical heavy nuclear targets. We investigate
the possibility of using liquid Helium-4 as a target material,
taking advantage of the favorable kinematic matching of the Helium
nucleus to light dark matter particles. Monte Carlo simulations are
performed to calculate the charge, scintillation, and triplet helium
molecule signals produced by recoil He ions, for a variety of
energies and electric fields. We show that excellent background
rejection can be achieved based on the ratios between different
signal channels. We also present some concepts for a liquid-helium-based dark matter detector. Key to the proposed approach is
the use of a large electric field to extract electrons from the
event site, and the amplification of this charge signal, through
proportional scintillation, liquid electroluminescence, or roton emission.
The sensitivity of the proposed detector to light dark matter
particles is estimated for various electric fields and light
collection efficiencies.
\end{abstract}

 \maketitle

\section{Introduction}
Dark matter, while evident on multiple astronomical length scales
through its gravitational effects, has an unknown intrinsic nature.
Data from primordial nucleosynthesis~\cite{Schramm-1998}, the cosmic
microwave background~\cite{Hu-2001}, structure formation~\cite{Tegmark-2004}, and microlensing
observations~\cite{Mellier-1999} imply that the dark matter cannot
be composed of baryons or active neutrinos, implying new physics
beyond the Standard Model. Experimental direct detection of dark
matter particles, illuminating their mass and interaction
properties, would therefore create crucial new scientific understanding
in both astrophysics and particle physics.

A particularly compelling model for dark matter is that it consists of Weakly Interacting Massive Particles, or WIMPs~\cite{Goodman-1985, JKG-1996}, with the feature that a massive particle in the early universe interacting through a weak-scale cross-section yields a thermal relic abundance approximately that observed for dark matter. Over the past few decades, models of WIMP dark matter
have centered on constrained minimal supersymmetry (CMSSM)
models~\cite{Giedt-2009}, which predict a stable neutralino with
mass greater than 40 GeV, limited to higher masses by the requisite
mass difference between the chargino and neutralino. Also, it is
commonly argued that in the context of supersymmetry it is most
natural for the dark matter mass to be comparable to the weak
scale~\cite{Goldberg-1983, Ellis-1984}. As a result, most direct
dark matter experiments have been designed to have excellent
sensitivity to dark matter particles with mass comparable to or greater than the weak scale, yet
most of these, including the CDMS \cite{CDMS-lite}, ZEPLIN
\cite{Akimov2010}, and XENON \cite{XENON10, Aprile2010} programs,
see no evidence for such high mass dark matter particles, down to
the recent XENON100 spin-independent cross-section limit of about
\(2 \times 10^{-45}~\mathrm{cm}^{2}\) at 55 GeV \cite{XENON100}. At the same time, the DAMA
\cite{Bernabei2010}, CoGeNT \cite{COGENT}, and CRESST
\cite{CRESST} experiments have seen event rate anomalies that can be
interpreted in terms of direct detection of
light WIMPs, and a number of astrophysical anomalies may be interpreted in terms of light WIMP annihilation\cite{Hooper2012}. Meanwhile, many new theories of light WIMPs have been developed, and this is currently an area of active development in particle phenomenology. Models
for light dark matter often involve a new mediator particle as well as the dark matter itself, and include the next to minimal supersymmetic model
(NMSSM) \cite{Gunion2006}, asymmetric dark matter
\cite{Nussinov1985}, WIMPless dark matter \cite{Feng2008}, singlet
scalars \cite{Burgess2001}, dark sectors with kinetic mixing
\cite{Pospelov2009}, mirror matter \cite{Foot2008}. These models can all evade constraints on light WIMPs from the cosmic microwave background \cite{Giesen}, the Large Hadron Collider \cite{Chatrchyan}, and Fermi-LAT \cite{Ackermann}.

Considerable excitement has been generated over the possibility that
dark matter particles are relatively low in mass. The difficulty is
detecting them, since lighter WIMPs have less kinetic energy and
only deposit a small fraction of it when elastically scattering with
standard heavy targets like germanium and xenon.

In general it is difficult for heavy
targets to be sensitive to light WIMPs, since for typical energy
thresholds they are only sensitive to a small part of the WIMP
velocity distribution.  Models of the WIMP velocity distribution typically assume a
Maxwellian distribution of \(f(v) = \exp{-(v+v_E)^2/v_0}^2\), where
$v_E\simeq244$~km/s is the velocity of the Earth around the Milky
Way, and $v_0\simeq230$~km/s is the virialized velocity of the
average particle that is gravitationally bound to the Milky
Way~\cite{Lewin-1996}. This distribution is expected to be roughly valid up to the Galactic
escape velocity $v_{esc}\simeq544$~km/s, above which the velocity
distribution is zero.
A plausible energy threshold for Xe, Ge, and He dark matter
experiments is about 5 keVr. But for a 5 GeV WIMP, such as predicted by asymmetric dark matter models \cite{Nussinov1985}, its velocity must
be particularly large to deposit at least 5 keV. This minimum
velocity, \(v_{min}\), is equal to \(v_{min} =
\sqrt{\frac{1}{2}\cdot{E_R}\cdot{M_T}}/r\), where \(E_R\) is the
recoil energy, $r$ is the WIMP-target reduced mass \(r=
M_D{\cdot}M_T/(M_D+M_T)\), \(M_D\) and \(M_T\) are the masses of the
dark matter particle and the mass of the target nucleus,
respectively. For \(E_R\) of 5 keV and \(M_D = 5~\mathrm{GeV}\),
\(v_{min}\) is equal to 1127, 864, and 427 km/s for Xe, Ge, and He
respectively. So for this example, \(v_{min}\) is above \(v_{esc}\) for Xe and
Ge, but not for He. The lower limit of the WIMP-target reduced
mass that a detector is sensitive to is given by
\begin{equation}
r_{limit} = \frac{1}{v_{esc}}{\cdot}\sqrt{E_t{\cdot}M_T/2},
\end{equation}
where \(E_t\) is the energy threshold. So a kinematic figure of merit for light WIMP detection is the
product of the energy threshold and the target mass, which should be minimized for the best light WIMP sensitivity.

This challenge of combined low energy threshold and low target mass
can likely be met through the use of liquid helium as a target material. In this
paper we investigate the use of liquid helium as a target for
light dark matter particles in the mass range of 1 to 10 GeV. In
Section II we outline the properties of liquid helium in the context
of particle detection, in Section III we describe possible
configurations of helium-based detectors. The detector can be
operated at $T\sim3$~K, adopting proportional scintillation or electroluminescence
for charge readout; or it can be operated at $T\sim100$~mK
using bolometers for light and charge readout. In Section IV we
examine the sensitivity of liquid helium detectors to light WIMPs. We
conclude in Section V.

\section{Liquid helium as a detector material}
Superfluid helium has been used for a detector material for many
applications. Most detector concepts take advantage of the special
excitations of the superfluid, and involve detection of phonons,
rotons, or quantum turbulence. One example is the HERON concept \cite{Lanou1987}
for pp-solar neutrino detection with rotons in superfluid
helium-4 at a temperature of ${\sim}100$ mK. The HERON researchers also considered using such an instrument to look for dark matter \cite{Lanou-1988, Adams-1996}, with the possibility that the roton/vortex generation by electrons in an applied electric field, combined with prompt roton detection, could be used for particle discrimination. Also, the roton signal could carry information about the nuclear recoil direction. Another is the ULTIMA concept \cite{Bradley1996} for dark matter detection with quantized turbulence in superfluid helium-3. Both of these concepts have been the subject of considerable
research and development in the past few decades.

Along with its many unusual properties related to superfluidity,
liquid helium also produces substantial scintillation light and
charge when exposed to ionizing radiation, just like liquid xenon
and liquid argon which are already used extensively in the search
for dark matter. Some ultracold neutron experiments already make use
of the prompt scintillation of liquid helium; for example the
measurement of the neutron beta-decay lifetime \cite{Doyle1994} and search
for the neutron permanent electric dipole moment at the Spallation Neutron Source
\cite{Golub1994,Beeketal}. The prompt scintillation yield in liquid helium is
well known, measured by the HERON collaboration to be about 20
photons/keV electron equivalent (keVee).

Depending on particle species, energetic particles elastic
scattering in helium can lead to electronic recoils (gamma ray, beta
scattering events) or nuclear recoils (neutron or WIMP dark matter
scattering events). The recoil electrons or He nucleus collide with
helium atoms, producing ionization and excitation of helium atoms
along their paths. The ionized electrons can be extracted by an
applied electric field. The decay of the helium excimers gives rise
to scintillation light. A fraction of the deposited energy is
converted into low-energy elementary excitations of the helium,
i.e., phonons and rotons. Signals from all these different channels
may in principle be used to detect and identify the scattering
events. The key for dark matter detection is to be able to suppress
the electronic recoils that make up most of the backgrounds from the
nuclear recoils that would make up a WIMP signal by use of event
discrimination. In this section, we shall estimate the nuclear and
electronic recoil signals due to ionization charge, prompt
scintillation light, metastable He$^{*}_{2}$ molecules. We shall
present results of Monte Carlo simulations showing that excellent
background rejection can be achieved for the purpose of WIMP dark
matter detection, based on the ratio between these different
signals.

\subsection{Low energy nuclear recoils in helium}
\subsubsection{Charge exchange processes}
\begin{figure*}[htb]
\includegraphics[scale=0.64]{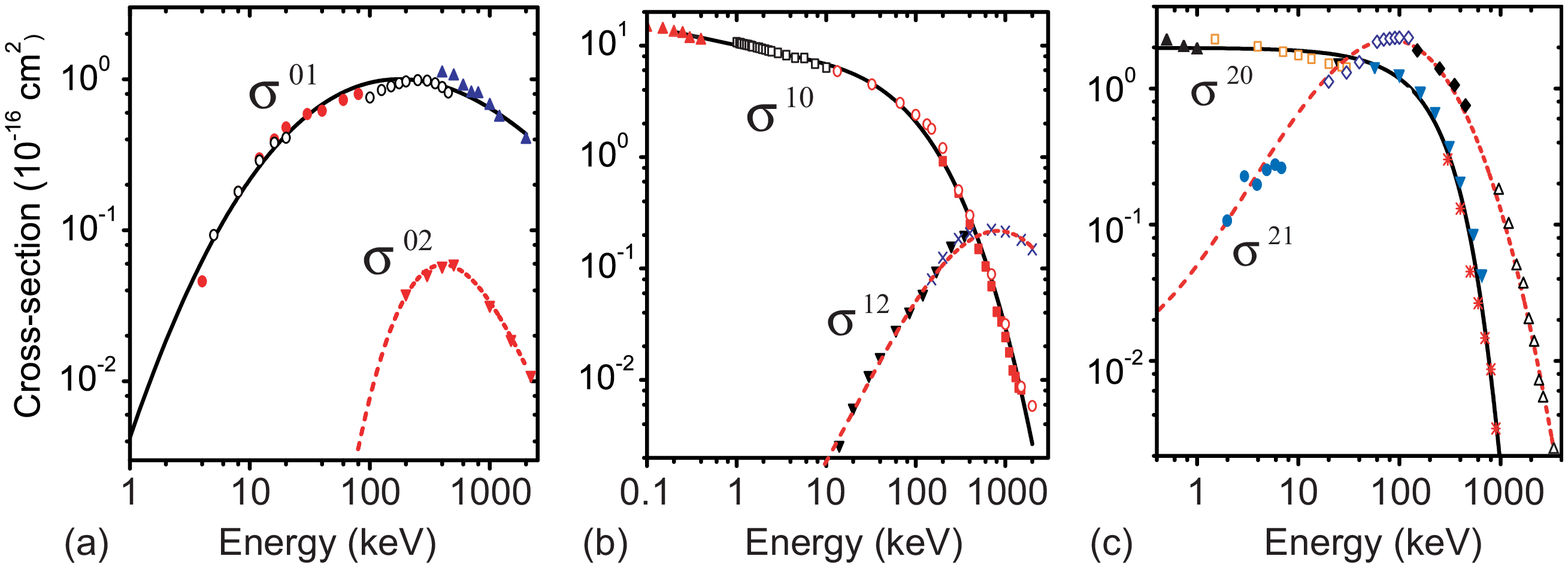}
\caption{(color online). Charge exchange cross sections due to
He$^{0}$, He$^{1+}$, and He$^{2+}$ interacting with ground state He
atoms. The curves were fitted to experimental data by polynomial
functions. (a) $\sigma^{01}$: --- (this work),
${\color{red}\bullet}$ (Ref.~\cite{Barnett1958}),
${\color{black}\circ}$ (Ref.~\cite{Rudnick1931}),
${\color{blue}\blacktriangle}$ (Ref.~\cite{Hvelplund1974});
$\sigma^{02}$: ${\color{red}- - -}$ (this work),
${\color{red}\blacktriangledown}$ (Ref.~\cite{Hvelplund1974}). (b)
$\sigma^{10}$: --- (this work), ${\color{red}\blacktriangle}$
(Ref.~\cite{Cramer1957}), $\Box$ (Ref.~\cite{Hegerberg1978}),
${\color{red}\circ}$ (Ref.~\cite{DuBois1989}),
${\color{red}\blacksquare}$ (Ref.~\cite{Pivovar1962});
$\sigma^{12}$: ${\color{red}- - -}$ (this work),
$\blacktriangledown$ (Ref.~\cite{Rudd1985}), ${\color{blue}\times}$
(Ref.~\cite{DuBois1989}). (c) $\sigma^{20}$:
--- (this work), $\blacktriangle$
(Ref.~\cite{Rivarola1979}), ${\color{red}\square}$
(Ref.~\cite{Grozdanov1980}), ${\color{blue}\blacktriangledown}$
(Ref.~\cite{Fulton1966}), ${\color{red}\ast}$
(Ref.~\cite{Pivovar1963}); $\sigma^{21}$: ${\color{red}- - -}$ (this
work), ${\color{blue}\bullet}$ (Ref.~\cite{Hertel1964}),
${\color{blue}\lozenge}$ (Ref.~\cite{Stich1985}), $\blacklozenge$
(Ref.~\cite{Allison1958}), $\vartriangle$
(Ref.~\cite{Wu1988}).}\label{Charge-exchange}
\end{figure*}
A WIMP dark matter scattering event in liquid helium would result a
recoil helium nuclei. Depending on the energy involved in the
scattering process, the recoil He can be a bare ion (He$^{2+}$) or a
dressed ion (He$^{1+}$), or even a neutral helium atom (He$^{0}$).
The recoil He dissipates its kinetic energy through collisions with
ground state He atoms. Such collisions can be elastic or inelastic
that lead to ionization or excitation of He atoms. The ionization
and excitation cross-sections are different for the recoil He ion in
different charge states. As the fast recoil He ion slows down,
interactions involving electron capture and loss by the projectile
become an increasingly important component of the energy loss
process. Charge transfer can produce residual ions without the
release of free electrons, and free electrons can be ejected from
the moving ion (or neutral) with no residual ions being formed.

Charge transfer cross sections are generally designated as
$\sigma^{if}$ where $i$ represents the initial charge state of the
moving ion, and $f$ is the charge state after the collision. For a
complete description of the full slowing down of a recoil He, we
need cross sections for one-electron capture $\sigma^{21}$ and
two-electron capture $\sigma^{20}$ for He$^{2+}$, one-electron
capture $\sigma^{10}$ and one-electron loss $\sigma^{12}$ for
He$^{1+}$, and one-electron loss $\sigma^{01}$ and two-electron loss
$\sigma^{02}$ for He$^{0}$. In Fig.~\ref{Charge-exchange}, we show
the six charge exchange cross-sections based on available
experimental data for He$^{0}$, He$^{1+}$ and He$^{2+}$. These cross
sections were least-squares fitted by simple polynomial functions of
the form
$\texttt{log}(\sigma^{if})=\sum_{n}{C_{n}(\texttt{log}~E)^{n}} $,
where the $C_{n}$'s are the fitting parameters, and $E$ is the
particle energy in keV. Smooth extrapolation was carried out where
the experimental data were lacking. Following the method by
Allison~\cite{Allison1958}, the fractions $F_{0}$, $F_{1}$, and
$F_{2}$ that the moving particle to be found in charge state 0, 1,
and 2 are given by
\begin{equation}
\begin{split}
&dF_{0}/dz=N[-F_{0}(\sigma^{01}+\sigma^{02})+F_{1}\sigma^{10}+F_{2}\sigma^{20}]\\
&dF_{1}/dz=N[-F_{1}(\sigma^{10}+\sigma^{12})+F_{0}\sigma^{01}+F_{2}\sigma^{21}]\\
&F_{2}=1-F_{0}-F_{1}
\end{split}
\label{Fraction}
\end{equation}
where $N\simeq2.2\times10^{22}$~cm$^{-3}$ is the number density of
liquid helium and $z$ is the path length along the particle track.
If the charge exchange cross-sections $\sigma^{if}$ do not vary as
the He ion moves, the equilibrium charge fractions $F^{\infty}_{0}$,
$F^{\infty}_{1}$, and $F^{\infty}_{2}$ as $z\rightarrow\infty$ are
given by Allision~\cite{Allison1958} as follows:
\begin{equation}
\begin{split}
&F^{\infty}_{0}=(f\sigma^{21}-a\sigma^{20})/D \\
&F^{\infty}_{1}=(b\sigma^{20}-g\sigma^{21})/D \\
&F^{\infty}_{2}=[(a-b)\sigma^{20}+g(a+\sigma^{21})-f(b+\sigma^{21})]/D
\end{split}
\label{Eq-fraction}
\end{equation}
in which
\begin{equation}
\begin{split}
&a=-(\sigma^{10}+\sigma^{12}+\sigma^{21}), ~ b=\sigma^{01}-\sigma^{21},\\
&f=\sigma^{10}-\sigma^{20}, ~ g=-(\sigma^{01}+\sigma^{02}+\sigma^{20}),\\
&D=ag-bf
\end{split}
\label{Parameter}
\end{equation}

\begin{figure}[htb]
\includegraphics[scale=0.36]{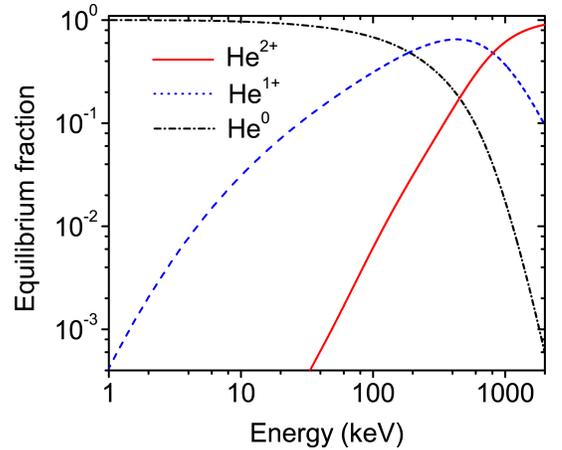}
\caption{(color online) Equilibrium fractions of the charge states of an energetic helium ion in liquid helium.}\label{Equilibrium fraction}
\end{figure}

\begin{figure}[htb]
\includegraphics[scale=0.4]{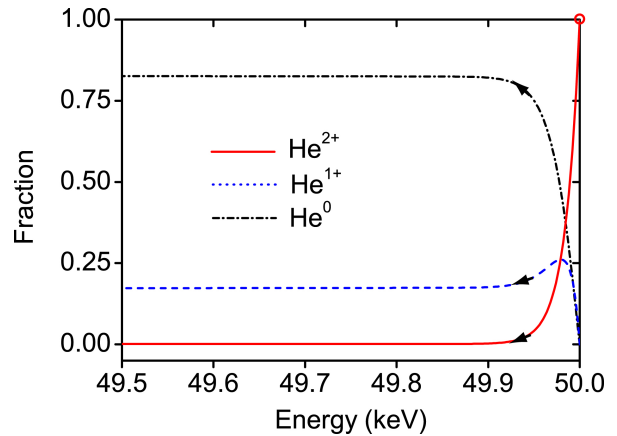}
\caption{(color online) Fractions of the charge states of an energetic helium ion as it slows down in liquid helium. The ion started as a He$^{2+}$ with initial energy of 50~keV, as indicated by the red circle. The arrows show how the fractions evolve as the particle loses its energy.}\label{Dynamic fraction}
\end{figure}

Fig.~\ref{Equilibrium fraction} shows the calculated equilibrium
charge fractions as a function of helium ion energy based on
Eq.~\ref{Eq-fraction} and Eq.~\ref{Parameter}. At energy higher than
a few thousands of keV, the helium ion appears primarily as a bare
ion He$^{2+}$, whereas in low energy regime ($<100$~keV) the
fraction of charge zero state He$^{0}$ dominates. These results are
derived based on the assumption that $\sigma^{if}$ does not vary as
the He ion moves. In reality, since the charge exchange
cross-sections depend on particle energy, as a He ion slows down in
liquid helium, the $\sigma^{if}$ in Eq.~\ref{Fraction} should change
as $z$ varies. In this situation, a full description of variation of
the charge fractions $F_{0}$, $F_{1}$, and $F_{2}$ is given by the
following equations
\begin{equation}
\resizebox{.8\hsize}{!}{$\begin{split}
&\frac{dF_{0}(E)}{dE}=\frac{N}{S(E)}\left[-F_{0}(\sigma^{01}+\sigma^{02})+F_{1}\sigma^{10}+F_{2}\sigma^{20}\right]\\
&\frac{dF_{1}(E)}{dE}=\frac{N}{S(E)}\left[-F_{1}(\sigma^{10}+\sigma^{12})+F_{0}\sigma^{01}+F_{2}\sigma^{21}\right]\\
&F_{2}(E)=1-F_{0}(E)-F_{1}(E)
\end{split}$}
\label{Fraction-1}
\end{equation}
where $S(E)$=$dE/dz$ is the total stopping power of a He ion in
liquid helium that describes the average energy loss of the He ion
per unit path length. $S(E)$ is the sum of the electronic stopping
power $S_{e}(E)$ (energy loss due to the inelastic collisions
between bound electrons in the medium and the ion) and the nuclear
stopping power $S_{n}(E)$ (energy loss due to the elastic collisions
between the helium atoms and the ion).
\begin{figure}[htb]
\includegraphics[scale=0.38]{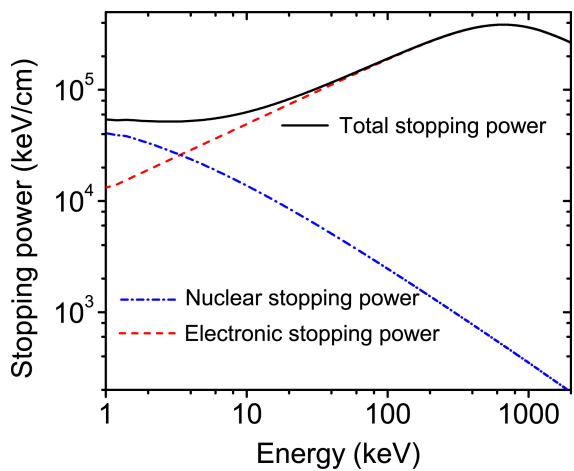}
\caption{(color online) Stopping power of a He ion in liquid
helium. Data are drawn from the National Institute of Standards
and Technology (NIST) database~\cite{NIST-stopping}.}\label{Stopping power}
\end{figure}
Fig.~\ref{Stopping power} shows the stopping power data drawn from
the National Institute of Standards and Technology (NIST)
database~\cite{NIST-stopping}. Knowing the stopping power $S(E)$ and
the charge exchange cross-sections $\sigma^{if}(E)$, one can
integrate Eq.~\ref{Fraction-1} to calculate the energy dependence of
the fractions of different charge states with a given initial
condition. An example is shown in Fig.~\ref{Dynamic fraction}.
We see that if we start with a bare ion He$^{2+}$ ($F_{2}=1$) at an
initial kinetic energy of 50~keV, as the ion slows down the
fractions of the different charge states $F_{0}$, $F_{1}$, and
$F_{2}$ quickly evolve to the equilibrium values. This is because
that due to the relatively large charge exchange cross-sections and
the high helium number density, many charge exchange collisions can
take place in a short path-length of the fast He ion. To achieve the
equilibrium charge fractions, only a few charge exchange collisions
are needed and the energy loss in this process is small. As a
consequence, we can safely use the equilibrium fractions of the
charge states to study the slowing down of a fast He ion in liquid
helium, with no need to consider the initial charge states.

\subsubsection{Ionization and excitation yields}
The ionization and excitation yields due to a recoil helium nuclei
moving in liquid helium are important premise parameters needed for
the design of a helium-based dark matter detector. Sato \emph{et
al.}~\cite{Sato1976} have studied the ionization and excitation
yields of an alpha particle (He$^{2+}$) in liquid helium using the
collision cross sections derived with the binary encounter
theory~\cite{Thomas1927}. In their analysis, the charge exchange
collisions are ignored and the fraction of the alpha particle energy
that is lost to elastic collisions with surrounding He atoms
(nuclear stopping) is not included. Nuclear stopping can become
dominant when the alpha particle energy is small, which is known as
the Lindhard effect~\cite{Lindhard}. The energy of a recoil helium
nuclei in a WIMP scattering event is expected to be relatively low
($\lesssim100$~keV). To obtain more reliable estimation of the
ionization and excitation yields from a recoil helium nuclei, we
present an analysis that systematically accounts for both the charge
exchange processes and the Lindhard effect.

\begin{figure*}[htb]
\includegraphics[scale=0.65]{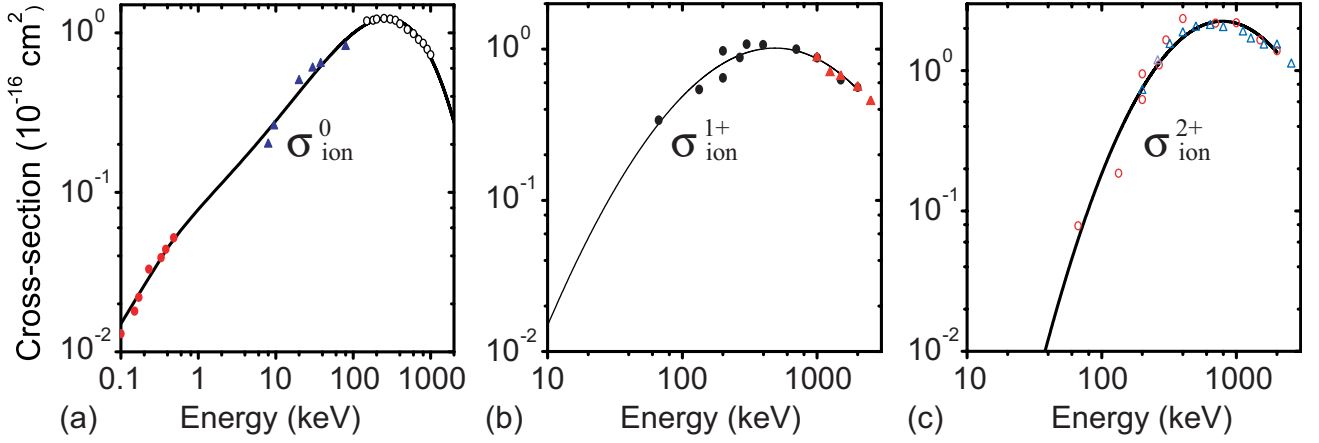}
\caption{(color online). Ionization cross-sections due to He$^{0}$,
He$^{1+}$, and He$^{2+}$ interacting with ground state He atoms. The
curves were fitted to experimental data by polynomial functions. (a)
$\sigma^{0}_{ion}$: --- (this work), ${\color{red}\bullet}$
(Ref.~\cite{Hayden1964}), ${\color{blue}\blacktriangle}$
(Ref.~\cite{Barnett1958-2}), $\circ$ (Ref.~\cite{Puckett1969}). (b)
$\sigma^{1+}_{ion}$: --- (this work), ${\bullet}$
(Ref.~\cite{DuBois1987}), ${\color{red}\blacktriangle}$
(Ref.~\cite{Santos2001}). (c) $\sigma^{2+}_{ion}$: --- (this work),
${\color{red}\circ}$ (Ref.~\cite{DuBois1987}),
${\color{blue}\vartriangle}$
(Ref.~\cite{Shah1985}).}\label{Ionization-cross-section}
\end{figure*}

\begin{figure}[htb]
\includegraphics[scale=0.45]{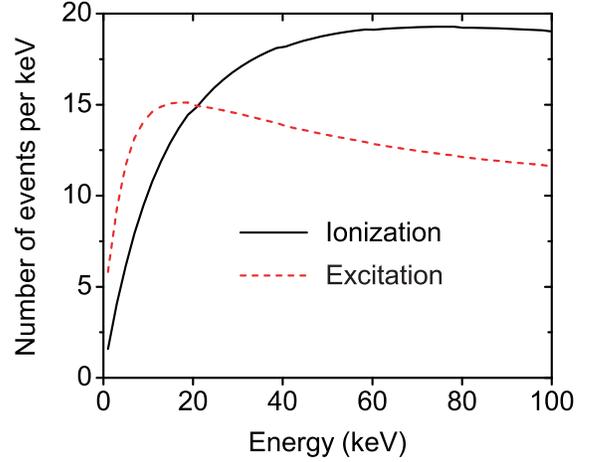}
\caption{(color online) Ionization and excitation yields of a recoil He ion in liquid helium as a function of the He ion energy.}\label{Ion-ex-yield}
\end{figure}

Assuming a continuous slowing down, the total number of free
electrons $N_{\texttt{el}}$ produced along the path of a recoil He
nuclei with an initial kinetic energy $E$ is given by
\begin{equation}
\resizebox{1.\hsize}{!}{$
\begin{split}
N_{\texttt{el}}&=N^{\texttt{Dir}}_{\texttt{el}}+N^{\texttt{Exc}}_{\texttt{el}}+N^{\texttt{Sec}}_{\texttt{el}}\\
&=\int^{E}_{0}\frac{NdE'}{S(E')}[F^{\infty}_{0}(E')\sigma^{0}_{\texttt{ion}}+F^{\infty}_{1+}(E')\sigma^{1+}_{\texttt{ion}}+F^{\infty}_{2+}(E')\sigma^{2+}_{\texttt{ion}}]\\
&+\int^{E}_{0}\frac{NdE'}{S(E')}[F^{\infty}_{0}(E')\left(\sigma^{01}+2\sigma^{02}\right)+F^{\infty}_{1+}(E')\sigma^{12}]\\
&+N^{\texttt{Sec}}_{\texttt{ion}} \label{Num-ionization}
\end{split}
$}
\end{equation}
Here $N^{\texttt{Dir}}_{\texttt{el}}$ and
$N^{\texttt{Exc}}_{\texttt{el}}$ are the number of electrons
produced in direct ionization and in charge exchange processes due
to He ion impact, and are given by the first and the second integral
terms on the right side of the equation.
$\sigma^{0}_{\texttt{ion}}$, $\sigma^{1+}_{\texttt{ion}}$, and
$\sigma^{2+}_{\texttt{ion}}$ are the direct ionization cross
sections due to He$^{0}$, He$^{1+}$, and He$^{2+}$ interacting with
ground state He atoms, respectively.
$N^{\texttt{Sec}}_{\texttt{el}}$ is the number of ionizations
produced by secondary electrons that have energy higher than the
ionization threshold of a He atom (24.6~eV). $F^{\infty}_{i}$
($i=0,1,2$) is the equilibrium fraction of charge state $i$ as given
by Eq.~\ref{Eq-fraction}. The ratio of
$N^{\texttt{Sec}}_{\texttt{el}}$ to $N_{\texttt{el}}$ decreases with
decreasing $E$ and is only a few percent when
$E\sim100$~keV~\cite{Sato1976}. We shall neglect
$N^{\texttt{Sec}}_{\texttt{el}}$ in the following analysis for
simplicity. To estimate the ionization yield, defined as
$Y_{\texttt{el}}=N_{\texttt{el}}/E$, the values of the direct
ionization cross sections are needed. In
Fig.~\ref{Ionization-cross-section} the experimental data for
$\sigma^{0}_{\texttt{ion}}$, $\sigma^{1+}_{\texttt{ion}}$, and
$\sigma^{2+}_{\texttt{ion}}$ are shown. We again fit the
experimental data by simple polynomial functions
$\texttt{log}(\sigma_{\texttt{ion}})=\sum_{n}{C'_{n}(\texttt{log}~E)^{n}}
$, and extrapolate the curves where the experimental data were
lacking. From Fig.~\ref{Equilibrium fraction} one can see that at
$E\lesssim100$~keV, the fraction of the charge zero state (He$^{0}$)
dominates. The available ionization and charge exchange cross
section data for He$^{0}$ in the energy range of 0.1$\sim$100~keV
allow us to make reliable fit and extrapolation for analyzing the
ionization yield. The calculated ionization yield $Y_{\texttt{el}}$
of a recoil He ion as a function of the ion energy is shown in
Fig.~\ref{Ion-ex-yield} as the black solid curve.

\begin{figure}[htb]
\includegraphics[scale=0.4]{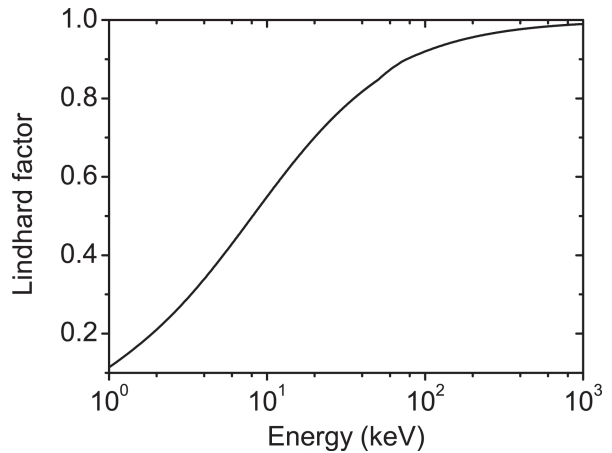}
\caption{Calculated Lindhard factor for a recoil He ion in liquid helium as a function of the He ion energy.}\label{Lindhard}
\end{figure}

The total number of excitations $N_{\texttt{ex}}$ produced by a
recoil He nuclei with an initial kinetic energy $E$ is given by
\begin{equation}
\begin{split}
N_{\texttt{ex}}=&\int^{E}_{0}\frac{NdE'}{S(E')}[F^{\infty}_{0}(E')\sigma^{0}_{\texttt{ex}}+F^{\infty}_{1}(E')\sigma^{1+}_{\texttt{ex}}\\
&+F^{\infty}_{2}(E')\sigma^{2+}_{\texttt{ex}}]+\tilde{N}_{\texttt{ex}} \label{Num-excitation}
\end{split}
\end{equation}
where $\sigma^{0}_{\texttt{ex}}$, $\sigma^{1+}_{\texttt{ex}}$, and
$\sigma^{2+}_{\texttt{ex}}$ are the total excitation cross sections
due to He$^{0}$, He$^{1+}$, and He$^{2+}$ interacting with ground
state He atoms, respectively. Here $\tilde{N}_{\texttt{ex}}$ is the
number of excitations produced by secondary electrons, which can
again be neglected at $E\lesssim100$~keV~\cite{Sato1976}.
Experimental excitation cross section data are limited. For
instance, Kempter \emph{et al.} estimated the excitation cross
sections due He atom impact, but only with collision energy below
600 eV~\cite{Kempter1975}; De Heer and Van Den Bos measured the
excitation cross sections for He$^{1+}$ incident on He, but only for
excitations to states with principle quantum number
$n\geqslant3$~\cite{Heer1965}. Instead of fitting the data to obtain
the excitation cross sections, we estimate the excitation yield
$Y_{\texttt{ex}}=N_{\texttt{ex}}/E$ based on the known electronic
stopping power as follows. The electronic stopping power $S_{e}(E)$
can be written as
\begin{equation}\label{E-stopping-power}
\resizebox{0.9\hsize}{!}{$\begin{split}
\frac{S_{e}}{N}&=F^{\infty}_{0}[\sigma^{0}_{\texttt{ion}}\left(Q_{\texttt{He}}+\bar{\varepsilon}_{0}\right)+\sigma^{0}_{\texttt{ex}}\Bar{Q}_{\texttt{ex}}+(\sigma^{01}+2\sigma^{02})(Q_{\texttt{He}}+{\lambda}E)]\\
&+F^{\infty}_{1}[\sigma^{1+}_{\texttt{ion}}\left(Q_{\texttt{He}}+\bar{\varepsilon}_{1}\right)+\sigma^{1+}_{\texttt{ex}}\Bar{Q}_{\texttt{ex}}+\sigma^{12}(Q_{\texttt{He}}+{\lambda}E)]\\
&+F^{\infty}_{2}[\sigma^{2+}_{\texttt{ion}}\left(Q_{\texttt{He}}+\bar{\varepsilon}_{2}\right)+\sigma^{2+}_{\texttt{ex}}\Bar{Q}_{\texttt{ex}}]
\end{split}$}
\end{equation}
Here $Q_{\texttt{He}}=24.6$~eV is the ionization energy of a helium
atom. $\bar{\varepsilon}$ is the average kinetic energy of secondary
electrons by He ion impact.
$\Bar{Q}_{\texttt{ex}}={\sum}Q_{ij}\sigma_{ij}/{\sum}\sigma_{ij}$ is
the mean excitation energy where $Q_{ij}$ and $\sigma_{ij}$ are the
He$(i$$\rightarrow$${j})$ excitation energy and the associated cross
section, respectively. Lack of detailed information, here we assume
$\Bar{Q}_{\texttt{ex}}$ to be the same for the incident He ion in
different charge states.
$\lambda=m_{\texttt{e}}/m_{\texttt{He}}\simeq1.36\times10^{-4}$
where $m_{\texttt{e}}$ and $m_{\texttt{He}}$ are the masses of an
electron and a He atom, respectively. In Eq.~\ref{E-stopping-power},
the energy transfer model is assumed such that in a charge-loss
collision, a stripped electron is ejected from the projectile with
nearly the same velocity as the projectile. Indeed the stripped
electrons are observed in the spectrum of secondary electrons
produced when He ion impacts on water vapor as a peak centered at
${\lambda}E$~\cite{Toburen1980}. An energy deposition of
$Q_{\texttt{He}}$+${\lambda}E$ is thus made when an electron is lost
from the projectile~\cite{Uehara2002}. In an electron capture
process, energy deposition is essentially due to the recoil of the
ionized He atom and is negligible. As a result, the terms in the
square brackets in Eq.~\ref{Num-excitation} can be derived based on
Eq.~\ref{E-stopping-power}
\begin{equation}\label{bracket-terms}
\resizebox{1.0\hsize}{!}{$
\begin{split}
&F^{\infty}_{0}\sigma^{0}_{\texttt{ex}}+F^{\infty}_{1}\sigma^{1+}_{\texttt{ex}}+F^{\infty}_{2}\sigma^{2+}_{\texttt{ex}}\\
&=\frac{1}{\Bar{Q}_{\texttt{ex}}}\{\frac{S_{e}}{N}-F^{\infty}_{0}[\sigma^{0}_{\texttt{ion}}\left(Q_{\texttt{He}}+\bar{\varepsilon}_{0}\right)+(\sigma^{01}+2\sigma^{02})(Q_{\texttt{He}}+{\lambda}E)]\\
&~~~-F^{\infty}_{1}[\sigma^{1+}_{\texttt{ion}}\left(Q_{\texttt{He}}+\bar{\varepsilon}_{1}\right)+\sigma^{12}(Q_{\texttt{He}}+{\lambda}E)]\\
&~~~-F^{\infty}_{2}[\sigma^{2+}_{\texttt{ion}}\left(Q_{\texttt{He}}+\bar{\varepsilon}_{2}\right)]\}
\end{split}
$}
\end{equation}
Plugging Eq.~\ref{bracket-terms} back into Eq.~\ref{Num-excitation},
the excitation yield can be derived as
\begin{equation}
\begin{split}
Y_{\texttt{ex}}&\simeq\frac{L}{\Bar{Q}_{\texttt{ex}}}-\frac{Q_{\texttt{He}}}{\Bar{Q}_{\texttt{ex}}}Y_{\texttt{el}}-\frac{1}{E}\int^{E}_{0}\frac{NdE'}{S(E')}\frac{1}{\Bar{Q}_{\texttt{ex}}}\cdot\\
&\{[F^{\infty}_{0}\sigma^{0}_{\texttt{ion}}\bar{\varepsilon}_{0}+F^{\infty}_{1}\sigma^{1+}_{\texttt{ion}}\bar{\varepsilon}_{1}+F^{\infty}_{2}\sigma^{2+}_{\texttt{ion}}\bar{\varepsilon}_{2}]\\
&+[F^{\infty}_{0}(\sigma^{01}+2\sigma^{02}){\lambda}E+F^{\infty}_{1}\sigma^{12}{\lambda}E]\}
\label{Excitation-yield}
\end{split}
\end{equation}
in which $L$ is the Lindhard factor, defined as
\begin{equation}
L=\frac{1}{E}\int^{E}_{0}\frac{S_{e}(E')dE'}{S(E')}.
\end{equation}
Lindhard factor designates the ratio of the energy given to the
electronic collisions to the total energy. A plot of the Lindhard
factor as a function of the recoil He ion energy is shown in
Fig.~\ref{Lindhard}. Since only the part of energy given to the
electronic collisions can be used as ionization and scintillation
signals, the Lindhard factor $L$ is important for the determination
of the sensitivity of WIMP detectors.

In order to calculate $Y_{\texttt{ex}}$ using
Eq.~\ref{Excitation-yield}, we need to make further approximations
on $\Bar{Q}_{\texttt{ex}}$ and $\bar{\varepsilon}$. Since the
dominant excitation process in low energy collisions between He
atoms and the projectile is He(1s$^{2}$$\rightarrow$1s2p) with an
excitation energy of 21.2~eV~\cite{Kempter1975}, we take
$\Bar{Q}_{\texttt{ex}}\simeq21$~eV for simplicity. The average
energy $\bar{\varepsilon}$ of the secondary electrons can be
expanded in power series of $E$. To the lowest order in $E$, we may
write $\bar{\varepsilon}\simeq{\gamma}(E-24.6$~eV) for $E>24.6$~eV.
Linear dependence of $\bar{\varepsilon}$ on $E$ is evidenced for
secondary electrons ejected by helium ion impact on water vapor with
energy $E\lesssim100$~keV~\cite{Uehara2002}. Furthermore, at small
$E$, $\bar{\varepsilon}$ is similar for He ion impact in different
charge states. We choose $\gamma=0.3$ for all charge states such
that the ratio between the calculated ionization yield
$Y_{\texttt{el}}$ and excitation yield $Y_{\texttt{ex}}$ agrees with
Sato \emph{et al}'s result at $E\sim100$~keV where the Lindhard
effect is mild. Note that variation of $\gamma$ does not affect
$Y_{\texttt{ex}}$ at small $E$. The calculated $Y_{\texttt{ex}}$ is
shown in Fig.~\ref{Ion-ex-yield} as the red dashed curve.

The drop of both the ionization yield and the excitation yield at
energies lower than about 50~keV is due to the drop of the
electronic collision cross sections in this energy regime, as well
as the loss of the He ion energy to elastic nuclear collisions
(Lindhard effect). As a comparison, for an energetic electron moving
in LHe, Sato \emph{et al}~\cite{Sato1976, Sato1974} estimated that
the total ionization yield and excitation yield are nearly constant
($Y^{(e)}_{el}\simeq22.7$~keV$^{-1}$ and
$Y^{(e)}_{ex}\simeq10.2$~keV$^{-1}$) in the energy range from a few
hundred keV down to about 1~keV.

\subsection{Signals in liquid helium}
\subsubsection{Charge signal}
Electrons and helium ions are produced along the track of an
energetic particle as a consequence of ionization or charge-exchange
collisions. Beside these processes, excited helium atoms produced by
the projectile with principal quantum number $n\geq3$ can autoionize
in liquid helium by the Hornbeck-Molnar process~\cite{Hornbeck}
\begin{equation}\label{H-M process}
\texttt{He}^{*}+\texttt{He}~\rightarrow~\texttt{He}^{+}_{2}+e^{-},
\end{equation}
since the 2~eV binding energy of He$^{+}_{2}$ is greater than the
energy to ionize a He($n\geq3$) atom. Based on the oscillator
strengths for the transitions between the ground state and the
various excited states of helium~\cite{Berkowitz1997}, slightly more
than one third of the atoms promoted to excited states will have a
principal quantum number of 3 or greater. All these electrons and
ions quickly thermalize with the liquid helium. The ions form helium
``snowballs" in a few picoseconds~\cite{Ovchinnikov}, and they do
not move appreciably from the sites where they are originated. On
the other hand, as the energy of the free electrons drops below
about 20 eV, the only process by which they can lose energy is
elastic scattering from helium atoms. Due to the low energy-transfer
efficiency (about $\lambda=1.36\times10^{-4}$ per collision), these
electrons make many collisions and undergo a random walk till their
energy drops below 0.1 eV, the energy thought to be necessary for
bubble state formation. Once thermalized, the electrons form bubbles
in the liquid typically within 4 ps~\cite{Hernandez}. Due to the
Coulomb attraction, electron bubbles and helium ion snowballs
recombine in a very short time and lead to the production of
He$^{*}_{2}$ excimer molecules
\begin{equation}\label{He-e reaction}
(\texttt{He}^{+}_{3})_{\texttt{snowball}}+(\texttt{e}^{-})_{\texttt{bubble}}~\rightarrow~\texttt{He}^{*}_{2}+\texttt{He}.
\end{equation}
When an external electric field is applied, some of the electrons
can escape the recombination and be extracted.

\begin{figure}[htb]
\includegraphics[scale=0.38]{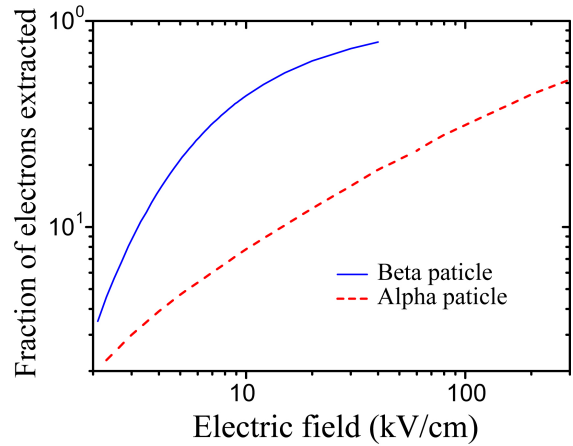}
\caption{(color online). Electron extraction fraction $q$ as a function of applied electric field. The blue solid curve represents the simulated electron extraction from beta tracks by Guo \emph{et al.}~\cite{Guo-2011}. The red dashed curved represents the simulated electron extraction from alpha tracks by Ito \emph{et al.}~\cite{Ito}.} \label{Charge-yield}
\end{figure}

At temperatures above 1~K, electron bubbles essentially move along
the electric field lines in the moving frame of the ions due to the
viscous damping~\cite{Guo-2009, Guo-2011}. In this situation, the
fraction $q$ of the electrons that can be extracted under an applied
field $\varepsilon$ depends largely on the initial electron-ion
separation and the ionization density along the projectile track.
The mean electron-ion separation has been determined to be about
60~nm for both beta particle ionization events~\cite{Guo-2011} and
alpha particle ionization events~\cite{Ito}. The energy deposition
rate for an electron of several hundred keV is approximately 50
eV/micron, whereas for an alpha particle of a few MeV the rate is 25
keV/micron~\cite{Adams}. The average energy needed to produce an
electron-ion pair has been measured to be about 42.3 eV for a beta
particle~\cite{Jesse} and about 43.3 eV~\cite{Ishida1992} for an
alpha particle. It follows that charge pairs are separated on
average about 850~nm along a beta particle track and only about
1.7~nm along the track of an energetic alpha particle. The
recombination along a beta particle track where the electron-ion
pairs are spatially separated is described by Onsager's geminate
recombination theory~\cite{Onsager}. For the highly ionizing track
of an alpha particle in liquid helium, the electrons feel the
attraction from all nearby ions and are harder to be extracted.
Jaffe's columnar theory of recombination is more applicable in this
situation~\cite{Jaffe, Kramers}. In Fig.~\ref{Charge-yield}, the
charge extraction from a beta particle track, simulated by Guo
\emph{et al.}~\cite{Guo-2011}, and that from an alpha particle
track, simulated by Ito \emph{et al.}~\cite{Ito}, are shown as the
blue solid curve and the red dashed curve, respectively. Note that in the low field regime, the measured charge collection by Ghosh~\cite{Ghosh-thesis} and Sethumadhavan~\cite{Sethumadhavan} for beta particles is higher than the predicted result by Guo
\emph{et al.}~\cite{Guo-2011}. Furthermore, these charge extraction analyses are for temperatures above 1~K. At very low temperatures, the ionized electrons can stray away from the field lines which enhances the charge extraction at a given applied field~\cite{Guo-2011}.

The mean electron-ion separation along the track of a low energy recoil He nuclei
should be similar to that for beta and alpha particles. The
ionization density along the He nuclei track can be estimated by
$(N_{el}+\frac{1}{3}N_{ex})/Z$, where $Z$=$\int^{E}_{0}dE'/S(E')$ is
the track length of the recoil He ion. Due to the Lindhard effect, a
major part of the projectile energy is lost to elastic collisions at
small $E$. Consequently the ionization density along the track of a
recoil He ion should be much lower than that along the track of an
energetic alpha particle. For instance, for a 10~keV recoil He
nuclei, the ions produced are on average separated by about 20~nm
along the track. At lower recoil energies, the separation between
ionizations becomes comparable or even larger than the mean
electron-ion separation. As a consequence, the charge extraction
fraction $q$ for low energy nuclear recoils is expected to be
similar to that for electron recoils. Due to the lack of
experimental information, in the following analysis we assume the
same $q$ for both low energy recoil He nucleus and beta particles.
The charge counts S2 for nuclear recoils and electron recoils are
thus given by $q(Y_{el}+\frac{1}{3}Y_{ex})E$ and
$q(Y^{(e)}_{el}+\frac{1}{3}Y^{(e)}_{ex})E$, respectively. Note that
the terms $\frac{1}{3}Y_{ex}$ and $\frac{1}{3}Y^{(e)}_{ex}$ account
for the ionizations produced by the auto-ionization of the excited
He atoms.

\subsubsection{Excitations and scintillation}
Excited helium atoms can be produced in excitation collisions. For
electron recoils, Sato \emph{et al.}~\cite{Sato1974} calculated that
83\% of the excited helium atoms produced in excitation collisions
are in the spin-singlet states and the rest 17\% are in triplet
states. For low energy nuclear recoils, however, the direct
excitations are nearly all to spin-singlet states~\cite{Sato1976,
Kempter1975}. This is because that since the total spin is
conserved, excitation to triplet states can occur only when both the
recoil He and the ground state He atom are excited simultaneously to
triplet states. This process requires more energy and has a lower
chance to occur. The excited atoms are then quickly quenched to
their lowest energy singlet and triplet states, He$^{*}(2^{1}$S) and
He$^{*}(2^{3}$S), and react with the ground state helium atoms of
the liquid, forming excited He$_{2}(A^{1}\Sigma_{u})$ and
He$_{2}(a^{3}\Sigma_{u})$ molecules
\begin{equation}\label{He-He reaction}
\texttt{He}^{*}+\texttt{He}~\rightarrow~\texttt{He}^{*}_{2}.
\end{equation}

He$^{*}_{2}$ excimer molecules are also produced as a consequence of
recombinations of electron-ion pairs. For geminate recombination,
experiments~\cite{Adams} indicate that roughly 50\% of the excimers
that form on recombination are in excited spin-singlet states and
50\% are in spin-triplet states. He$^{*}_{2}$ molecules in highly
excited singlet states can rapidly cascade to the first-excited
state, He$_{2}(A^{1}\Sigma_{u})$, and from there radiatively decay
in less than 10 ns to the ground state~\cite{McKinsey-thesis},
He$_{2}(X^{1}\Sigma_{g})$, emitting ultraviolet photons in a band
from 13 to 20~eV and centered at 16~eV. As a consequence, an intense
prompt pulse of extreme-ultraviolet (EUV) scintillation light is
released following an ionizing radiation event. These photons can
pass through bulk helium and be detected since there is no
absorption in helium below 20.6 eV.

The radiative decay of the triplet molecules
He$_{2}(a^{3}\Sigma_{u})$ to the singlet ground state
He$_{2}(X^{1}\Sigma_{g})$ is forbidden since the transition involves
a spin flip. The radiative lifetime of an isolated triplet molecule
He$_{2}(a^{3}\Sigma_{u})$ has been measured in liquid helium to be
around 13~s~\cite{McKinsey-1999}. The triplet molecules, resulting
from both electron-ion recombination and from reaction of excited
triplet atoms, diffuse out of the ionization track. They may
radiatively decay, react with each other via bimolecular Penning
ionization~\cite{Keto-1974-1}, or be quenched at the container
walls. Experimentally, these molecules can be driven by a heat
current to quench on a metal detector surface and be detected as a
charge signal~\cite{Mehrotra-1979, Zmeev-2012}.

Note that non-radiative destruction of singlet excimers by the bimolecular Penning ionization process can lead to the quenching of the prompt scintillation light. Such quenching effect has been observed for energetic alpha particles in liquid helium \cite{Adams, Roberts-1973}. At temperatures above 2~K, the singlet excimers can diffuse on the order of 10 nm in their 10~ns lifetime \cite{McKinsey-2005}. Based on the discussion presented in the previous section, the mean separation of the excimers along the track of a low energy recoil helium can be greater than the diffusion range of the singlet excimers. The quenching of the prompt scintillation for low energy nuclear recoils should thus be small. At low temperatures, the quenching effect may be significant. However, it has been observed that even for the highly ionizing track of an energetic alpha particle, the light quenching becomes mild below about 0.6~K~\cite{Roberts-1973}. This is presumably due to the trapping of the excimers on quantized vortex lines that are created accompanying the energy deposition of the recoil helium~\cite{Zmeev-2012}. Such trapping limits the motion of the excimers and hence reduces the light quenching. Lack of experimental knowledge about the decay rates at which bimolecular Penning processes occur among the different excimers, we shall not include the quenching effect in our analysis.

Based on the above discussion, the prompt scintillation photons
($S1^{(e)}$) and triplet molecules ($S3^{(e)}$) produced by an
electron recoil event are given by
\begin{equation}
\begin{split}
S1^{(e)}=&E\cdot\left[Y^{(e)}_{el}\cdot(1-q)\cdot50\%+Y^{(e)}_{ex}\cdot86\%\cdot\frac{2}{3}\right.\\
&\left.+Y^{(e)}_{ex}\cdot86\%\cdot\frac{1}{3}\cdot(1-q)\right]\label{S1-e-count}
\end{split}
\end{equation}
\begin{equation}
\begin{split}
S3^{(e)}=&E\cdot\left[Y^{(e)}_{el}\cdot(1-q)\cdot50\%+Y^{(e)}_{ex}\cdot14\%\cdot\frac{2}{3}\right.\\
&\left.+Y^{(e)}_{ex}\cdot14\%\cdot(1-q)\cdot\frac{1}{3}\right]\label{S3-e-count}
\end{split}
\end{equation}
The factor $2/3$ accounts for the fraction of the excited atoms that
do not undergo autoionization. The above two formulas assume that
the recombination of electron-ion pairs produced by the
autoionization of singlet (or triplet) helium atoms tends to
generate only singlet (or triplet) helium excimers. The
justification for this assumption is that the energy of the
electrons produced in the Hornbeck-Molnar process is low (less than
2 eV). These electrons do not move very far from their parent ions,
hence their spin correlation with their parent ions is likely strong
enough to survive the whole ionization-recombination process. As for
the nuclear recoils, the $S1^{(n)}$ and $S3^{(n)}$ counts are given
by
\begin{equation}
\begin{split}
S1^{(n)}=&E\cdot\left[Y_{el}\cdot(1-q)\cdot50\%+Y_{ex}\cdot\frac{2}{3}\right.\\
&\left.+Y_{ex}\cdot\frac{1}{3}\cdot(1-q)\right]
\label{S1-n-count}
\end{split}
\end{equation}
\begin{equation}
S3^{(n)}=E{\cdot}Y_{el}\cdot(1-q)\cdot50\%
\label{S3-n-count}
\end{equation}
Since the excited atoms are assumed to be all in singlet states for
nuclear recoils, the triplet molecules are generated solely as a
consequence of the recombination of charge pairs produced in direct
ionization collisions.

For the readers' convenience, in table~\ref{Table-1}, we summarize
the formulas that we used to estimate the S1, S2, and S3 counts for
both nuclear recoils and electron recoils with incident energy of
$E$.

\begin{table*}[htb]
\caption{The yields of prompt scintillation (S1), charge (S2), and
He$^{*}_{2}$ triplet molecules (S3) for nuclear recoils and electron
recoils with incident energy of $E$ in liquid helium.}
\label{Table-1}
\begin{ruledtabular}
\begin{tabular}{c|c|c}
     & Nuclear recoils & Electron recoils\\
  \hline
  S1 & $E\cdot[0.5{\cdot}Y_{el}\cdot(1-q)+0.67{\cdot}Y_{ex}+0.33{\cdot}Y_{ex}\cdot(1-q)]$ & $E\cdot[0.5{\cdot}Y^{(e)}_{el}\cdot(1-q)+0.57{\cdot}Y^{(e)}_{ex}+0.29{\cdot}Y^{(e)}_{ex}\cdot(1-q)]$ \\
  \hline
  S2 & $E{\cdot}(Y_{el}+0.33{\cdot}Y_{ex}){\cdot}q$ & $E{\cdot}(Y^{(e)}_{el}+0.33{\cdot}Y^{(e)}_{ex}){\cdot}q$ \\
  \hline
  S3 & $E{\cdot}Y_{el}{\cdot}0.5\cdot(1-q)$ & $E\cdot[0.5{\cdot}Y^{(e)}_{el}\cdot(1-q)+0.093{\cdot}Y^{(e)}_{ex}+0.047{\cdot}Y^{(e)}_{ex}\cdot(1-q)]$ \\
\end{tabular}
\end{ruledtabular}
\end{table*}

\subsection{Discrimination of nuclear recoil and electronic recoil}

\subsubsection{Ratios of the signals from different channels}
\begin{figure}[htb]
\includegraphics[scale=0.4]{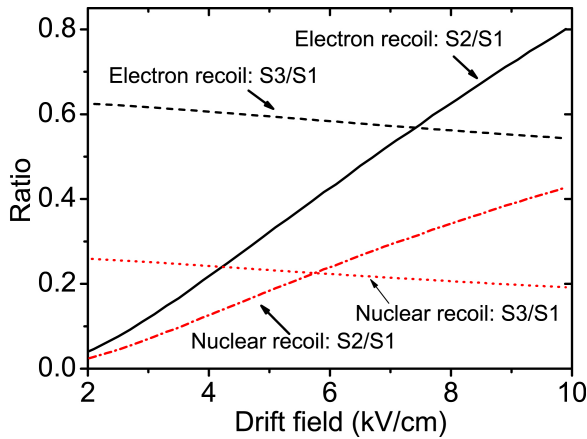}
\caption{(color online) The ratio of the counts from different signal channels for 10 keV nuclear recoil and electronic recoil events as a function of the applied electric field.}
\label{ratio-1}
\end{figure}

The success of a direct dark matter experiment will depend in its
ability to distinguish between electron recoils and nuclear recoils.
Discrimination between both types of recoils can be done by looking
at the ratio of the counts from different signal channels. These
ratios depend on the event type, the recoil energy, and the applied
electric field. The formulas listed in table~\ref{Table-1} allow us
to estimate these ratios. As an example, in Fig.~\ref{ratio-1}, the
calculated ratios of S2/S1 and S3/S1 are shown as a function of the
applied electric field for both the electron recoils and nuclear
recoils with a recoil energy of 10~keV. The S2/S1 ratio for both
electron recoils and nuclear recoils increases with the applied
electric field. This is because at higher fields more electrons can
be extracted, which enhances the S2 counts and at the meanwhile
reduces the S1 counts since less electrons recombine with the ions
to form singlet molecules. The difference of the S2/S1 ratio between
electron recoils and nuclear recoils becomes greater at higher
fields, which means that better discrimination based on S2/S1 can be
achieved at higher fields.

\begin{figure}[htb]
\includegraphics[scale=0.4]{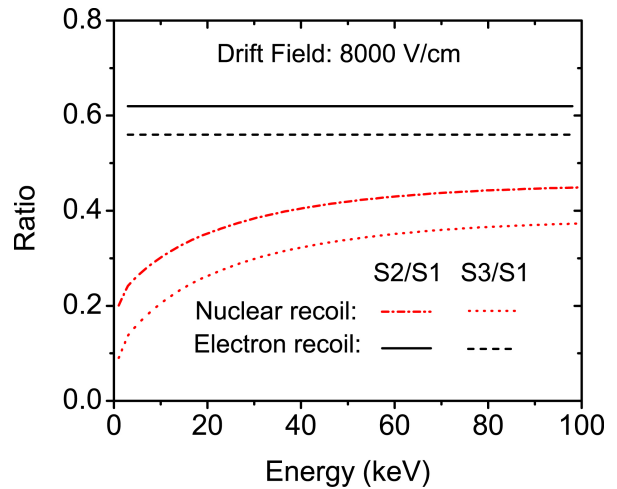}
\caption{(color online) The ratio of the counts between different signal channels for nuclear recoil and electronic recoil events as a function of the event energy. The applied electric field is 8 kV/cm.}
\label{ratio-2}
\end{figure}

In Fig.~\ref{ratio-2}, we show the calculated ratios of S2/S1 and
S3/S1 as a function of the event energy for both electron recoils
and nuclear recoils at an applied field of 8~kV/cm. Since we take
the ionization and excitation yields for electron recoils to be
constants, the S2/S1 and S3/S1 ratios for electrons recoils are
independent of the recoil energy. For nuclear recoils, both the
S2/S1 and S3/S2 ratios decrease with decreasing recoil energy. Note
that the ratios evaluated here are based on the calculated average
counts from the different signal channels. In real experiment, there
always exist number uncertainties of the counts. At low recoil
energies where the counts are small, the relative uncertainties of
the counts as well as the ratios of the counts between different
channels become large, which limits the discrimination of the two
types of recoils. For helium detector, as we can see from
Fig.~\ref{ratio-2}, the S2/S1 and S3/S1 curves for nuclear recoils
bend away from those for electron recoils, which compensates the
effect due to count uncertainty. As we shall show later, excellent
event discrimination can still be achieved even down to a few keV
energy regime.

\subsubsection{Scintillation efficiency factor}
The quantities that can be measured experimentally for a recoil
event are the counts from the different signal channels. One can
plot, for instance, the S2/S1 ratio against the S1 counts. However,
the conversion between S1 counts to the event energy is different
for electron recoils and nuclear recoils. For electron recoils, the
event energy is proportional to the S1 counts, since the ionization
and the excitation yields are taken to be constant. For nuclear
recoils, such conversion is nonlinear. The effective scintillation efficiency  $L_{eff}$ describes the difference between the amount of
energy measured in the detector between both types of recoils. In
the notation used in the field, the keV electron equivalent scale
(keVee) is used to quantify a measured signal in terms of the energy
of an electron recoil that would be required to generate it.
Similarly the keVr scale is used for nuclear recoil events. For a
nuclear recoil of energy $E_{r}$, the electron recoil event that
would produce an equivalent S1 signal is given by
\begin{equation}
E_{e}(keVee) = L_{eff}{\times}E_{r}(keVr).
\label{Leff-eq}
\end{equation}

\begin{figure}[htb]
\includegraphics[scale=0.38]{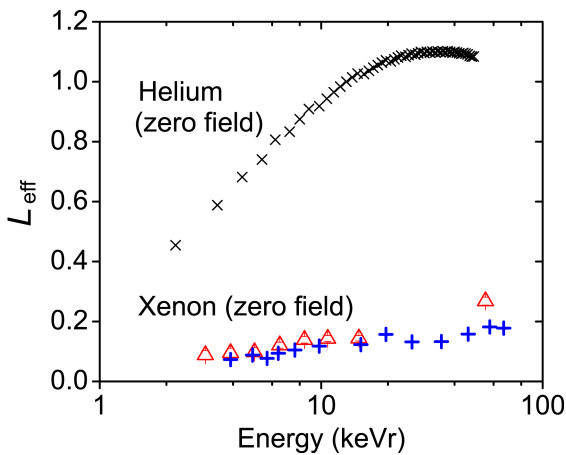}
\caption{(color online) The effective quenching factor ${L}_{eff}$ as a function of the recoil event energy. The ${\color{black}\times}$ represents the calculated ${L}_{eff}$ for helium under zero applied electric field. The measured data for liquid Xenon by G. Plante \emph{et al.}~\cite{Plante-2011} (${\color{red}\triangle}$) and by A. Manzur \emph{et al.}~\cite{Manzur-2010} (${\color{blue}+}$) are also shown.}
\label{Leff}
\end{figure}

By definition $L_{eff}$ is the nuclear recoil scintillation
efficiency relative to that of an electron recoil of the same energy
at zero field. Experimentally, the conversion between S1 and the
electron equivalent scale keVee can be established using gamma line
sources, for example the $^{57}$Co 122 keV line. The nuclear recoil
response as a function of energy can be established using neutron
scattering, either in a mono-energetic neutron scattering
experiment, or by using a neutron source with a broad energy
distribution and comparing the observed shape of the nuclear recoil
spectrum with detailed Monte Carlo simulations. Using the formulas
listed in Table~\ref{Table-1}, we can estimate the $L_{eff}$ by
calculating the ratio of the energies of the two types of recoil
events that give the same S1 counts. The result is shown in
Fig.~\ref{Leff}.

The event discrimination ability of a detector drops drastically
below a certain threshold S1 counts. For a given energy threshold in
keVee scale, a detector with a higher $L_{eff}$ has lower nuclear
recoil energy threshold, hence would be sensitive to low energy
WIMPs. In Fig.~\ref{Leff}, we also show the experimentally measured
$L_{eff}$ data for liquid Xenon \cite{Plante-2011,Manzur-2010}. In
the low energy regime of a few keV, Xenon-based detector only has a
$L_{eff}$ of less than 0.1 while helium detector has a $L_{eff}$
above 0.4. So while liquid helium has substantially lower scintillation yield for electron recoils, this is unlikely to be the case for nuclear recoils.

\subsubsection{Rejection power}
\begin{figure}[htb]
\includegraphics[scale=0.4]{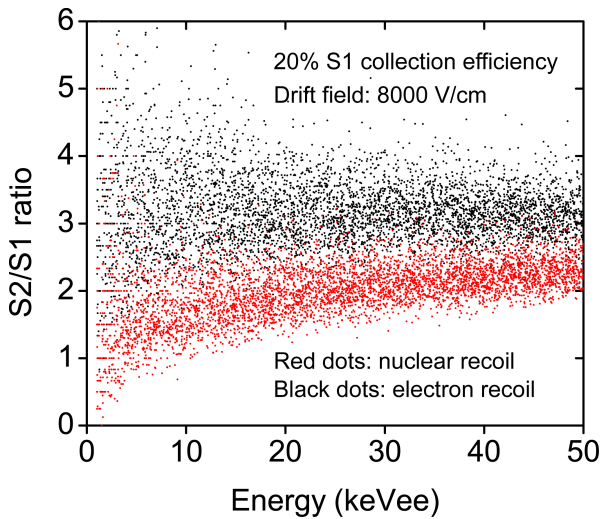}
\caption{(color online) Monte Carlo simulation of S1/S2 ratio for nuclear recoils (red dots) and electronic recoils (black dots). The S1 scintillation light collection efficiency is assumed to be 20\%. The applied electric field is 8~kV/cm. The event energy is in the keV electron equivalent energy scale (keVee).}
\label{Scatter-plot}
\end{figure}

The uncertainty of the signal counts limits the discrimination
between the nuclear recoils and the electron recoils at low
energies. To study this effect, we performed a Monte Carlo
simulation. For a given recoil energy $E$ in electron equivalent
energy scale, we randomly generate S1 and S2 counts for a nuclear
recoil event and an electron recoil event, assuming a Poisson
distribution of the counts with mean count values given by the
formulas listed in Table~\ref{Table-1}. In each trial, the ratios of
S2/S1 for a nuclear recoil and for an electronic recoil are
evaluated and represented by a red dot and a black dot in the S2/S1
versus $E$ plot. $10^7$ trials are carried out for each energy. An
example is shown in Fig.~\ref{Scatter-plot}. To match with real
experiments, we assume that only 20\% of the scintillation photons
(S1) are collected (typical for a two-phase detector as we shall
discuss later), and that all the extracted electrons (S2) under the
applied drift electric field can be detected. A clear
separation can be seen between the electron recoil band and the
nuclear recoil band, a necessary criterion for any direct dark
matter experiment. At low energies, the two bands overlap due to the
large scattering of the S2/S1 value. This large scattering is caused
by the large relative uncertainty of the counts when their mean
values of the Poisson distributions are small.

\begin{figure}[htb]
\includegraphics[scale=0.55]{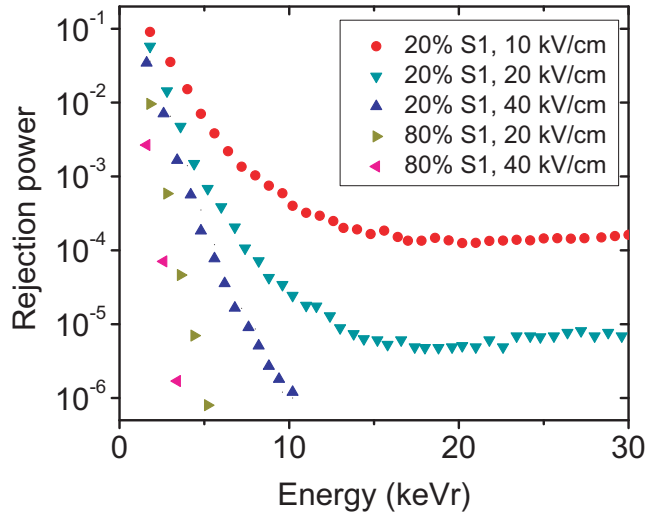}
\caption{(color online) Calculated rejection power for a helium detector as a function of event energy in keVr scale.}
\label{Discrimination}
\end{figure}

To calculate the rejection power, we divide the two bands in energy
slices. We select the lower half of the nuclear recoil band as our
WIMP region of interest. At low energies it is crucial to know what
percentage of the electronic recoil band leak into the lower 50\%
nuclear recoil band. The rejection power (or sometimes called
discrimination power) is found as the fraction of electronic recoil
events below the nuclear recoil centroid. A full description on this
method has been given by A. Manalaysay~\cite{Manalaysay-thesis}. In
Fig~\ref{Discrimination}, we show the calculated rejection power as
a function of event energy in keVr scale at several applied electric
fields and with different S1 collection efficiencies. At low
energies, the ability to distinguish electron and nuclear recoils is
degraded because of the lack of charge and light signal. But above a
few keV, discrimination power is predicted to improve considerably,
and this should allow for low-background operation and a sensitive
WIMP search. The discrimination is better at higher fields or with
higher S1 collection efficiency. We considered fields up to
40~kV/cm. It has been shown that such high fields can be readily
applied in liquid helium~\cite{Ito}. Indeed, the design electric field value of the Spallation Neutron Source (SNS) neutron EDM experiment is 50 kV/cm \cite{Long-2006, Ito-2007, Beck-2011}. As we shall discuss later, for
a single-phase helium detector operated at very low temperatures,
sensitive bolometers immersed in liquid helium may be used to read
out the light and the charge signals. In this case, 80\% S1
collection may be possible with the detector inner surface being
covered by bolometer arrays. Note that the rejection power analysis is based on the the charge extraction curve shown in Fig.~\ref{Charge-yield}. The actual charge extraction at a given field could be higher, especially at low temperatures where the ionized electrons can stray away from field lines~\cite{Guo-2011}. Considering this factor, the actual rejection power could be better than the results shown in Fig~\ref{Discrimination}.

\section{Helium-based detector}
For dark matter detection, we propose to detect both prompt
scintillation and charge in liquid helium-4, using a time
projection chamber design. This is essentially the same approach
used in Ar and Xe detectors~\cite{Aprile-2004, Benetti, Bolozdynya},
which has proven to be very effective, providing excellent position
resolution and gamma ray discrimination. Based on the readout
schemes for the light and charge signals, we discuss two proposals
for a liquid-helium-based dark matter detector. One proposal is
to operate the detector at high temperature regime ($\sim3$~K) using
photomultiplier tube arrays for signal readout, and the second
proposal is to operate the detector at low temperatures
($\sim100$~mK) using bolometer arrays for signal readout.

\subsection{High temperature operation scheme}
Operating a helium detector at relatively high temperatures is
favored in economy since such a detector does not require
complicated large-scale dilution refrigeration system. At high
temperatures where the helium vapor density is high, some
existing technologies for charge signal amplification may be ready
applied to the helium detector, such as the proportional
scintillation in a two-phase chamber that has been used in Argon,
Krypton, and Xenon detectors~\cite{Conde-1967, Aprile-2004, Benetti,
Bolozdynya}, or the electron avalanche in a Gas Electron Multiplier
(GEM)~\cite{Buzulutskov-2005,Buzulutskov-2012, Breskin}.

\begin{figure}[htb]
\includegraphics[scale=0.55]{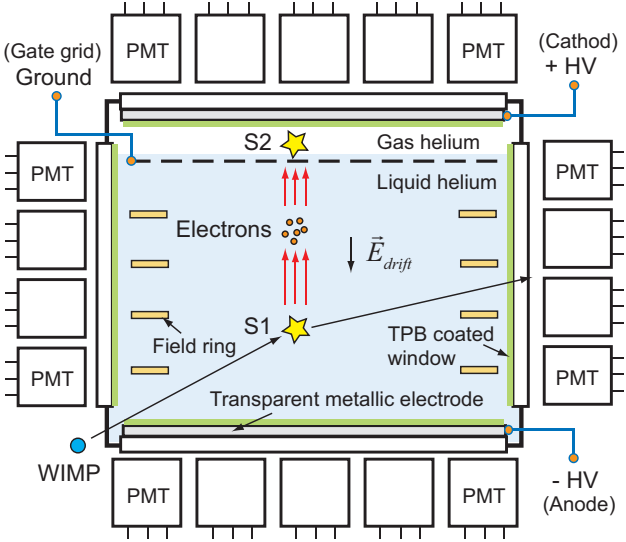}
\caption{(color online). A schematic of a two-phase helium-based dark matter detector.} \label{HighT-setup}
\end{figure}
A conceptual schematic of a two-phase helium-based time projection
chamber is shown in Fig.~\ref{HighT-setup}. Ionizing radiation events
in liquid helium produce both prompt scintillation light (S1) and
ionizations. A drift electric field can be maintained between the anode and the cathode. Some ionized electrons can be extracted into the gas phase and caused to emit 16~eV EUV photons (S2) by a strong field in the vapor. The anode and cathode may be a transparent material coated with Indium Tin Oxide such as recently demonstrated in the DarkSide-10 experiment \cite{Darkside}, so produce a uniform proportional scintillation field and eliminate liquid helium scattering events below the cathode. To detect the EUV photons, as is typical for scintillation detection in liquid argon, liquid neon,
and liquid helium, the inner surface of the chamber windows and the transparent electrodes can be
coated with tetraphenyl butadiene (TPB) wavelength shifter. The EUV
scintillation light is converted to the visible on the TPB coating,
which has approximately 100\% photon-to-photon conversion efficiency
at the helium scintillation wavelength of 80 nm \cite{Doyle1994}.
Photomultiplier tube (PMT) array placed outside the helium chamber
can be used to detect the converted photons. S1 light detection efficiency of about 20\% may be expected in such a detector, similar to that measured by the DarkSide collaboration \cite{Darkside}, which recently demonstrated a zero-field signal yield of 9.1 photoelectrons/keVee in a two-phase liquid argon detector. With time projection readout, the time between the S1 and S2 signals indicates the depth
($z$) of the event, while the hit pattern in the top array of
photomultipliers gives the $x-y$ position of the event. This allows
determination of a low-background central fiducial volume, which is
used for the dark matter search. Events close to chamber surface may
be discarded, and the S2/S1 ratio provides discrimination power
against gamma ray background. Any gamma ray or neutron events that
cause multiple scatters will generate multiple S2 signals, and these
may also be discarded.

\begin{figure}[htb]
\includegraphics[scale=0.58]{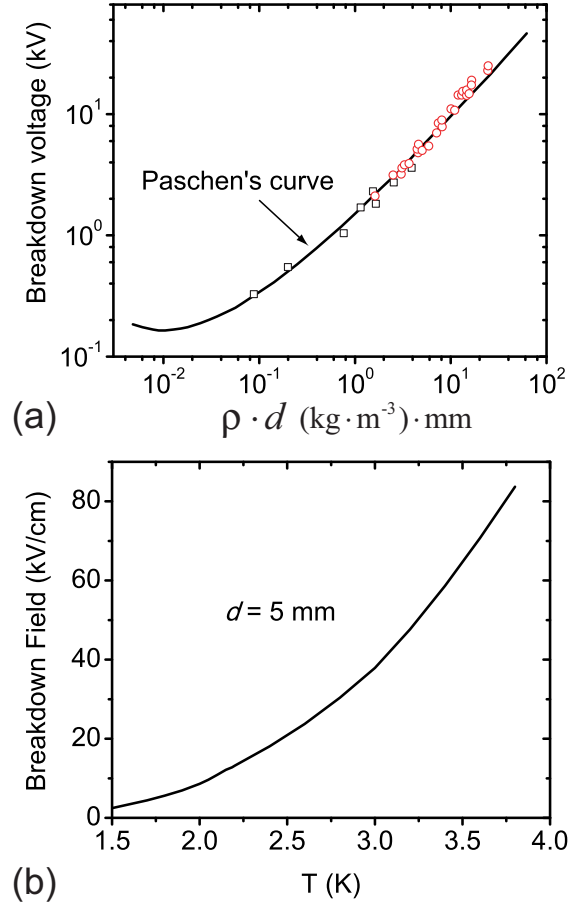}
\caption{(color online). (a)~The dielectric breakdown voltage in helium gas as a function of the product of the gas density $\rho$ and the electrode separation $d$. The red circles (${\color{red}\circ})$ and the black squares (${\square}$) are experimental data from Ref.~\cite{Gerhold-1972} and Ref.~\cite{Satow-1998}, respectively. The solid curve represents the Paschen's curve obtained by fitting the experimental data~\cite{Satow-1998, Winkelnkemper-1977}. (b)~The dielectric breakdown field for saturated helium vapor as a function of temperature.} \label{Breakdown field}
\end{figure}

The extracted electrons moving in helium vapor undergo collisions with helium atoms. Due to their very small mass the electrons can give up almost no energy to
the helium atoms in the course of elastic collisions. When the kinetic energy of the electrons builds up over a few mean free pathes to exceed the excitation threshold ($\sim20$~eV) of helium atoms, inelastic collisions between the electrons and the helium atoms, which lead to the production of excited helium atoms, can occur. The excited helium atoms in spin-singlet states can react with surrounding helium atoms and decay to the ground state by emitting 16 eV EUV photons (S2). The S2 strength increases with the applied field. However, when the electric field in vapor is too strong, the electron energy can exceed the ionization threshold ($\sim24.6$~eV) of helium atoms. In this situation, charge multiplication in gas occurs, and eventually avalanche breakdown can take place. Given the premium on high electric field for getting good event discrimination, it is advantageous to operate the detector at a field in the vapor only slightly below the breakdown field. The breakdown voltage $V_{bd}$ of helium gas in a uniform field generated by a pair of electrodes separated by a distance $d$ is given by the Paschen's law~\cite{Paschen-1889}
\begin{equation}
V_{bd}=\frac{A{\cdot}\rho{\cdot}d}{\texttt{ln}(\rho{\cdot}d)+B}
\label{V_breakdown}
\end{equation}
where $\rho$ is gas density. $A$ and $B$ are experimentally determined constants. Some available experimental data of the breakdown voltage for helium gas are shown in Fig.~\ref{Breakdown field} (a) \cite{Gerhold-1972, Satow-1998}. The solid curve represents a typical Paschen's curve for helium gas obtained by fitting the experimental data using Eq.~\ref{V_breakdown} \cite{Satow-1998, Winkelnkemper-1977}.

For saturated helium vapor in a two-phase helium detector, the gas density as a function of temperature is known~\cite{Sciver-book}. We can thus use the Paschen's curve to calculate the breakdown field for a two-phase helium detector as a function of the operation temperature for a given electrode separation $d$ in the vapor. In Fig.~\ref{Breakdown field} (b), the calculated breakdown field at $d=5$~mm is shown. As we can see, the breakdown field increases drastically with temperature due to the increased vapor density. At 3.2~K, the breakdown field is about 40~kV/cm. Under a drift field $E_{\texttt{drift}}$ that is close to this breakdown field, good event discrimination power is expected.

\begin{figure}[htb]
\includegraphics[scale=0.33]{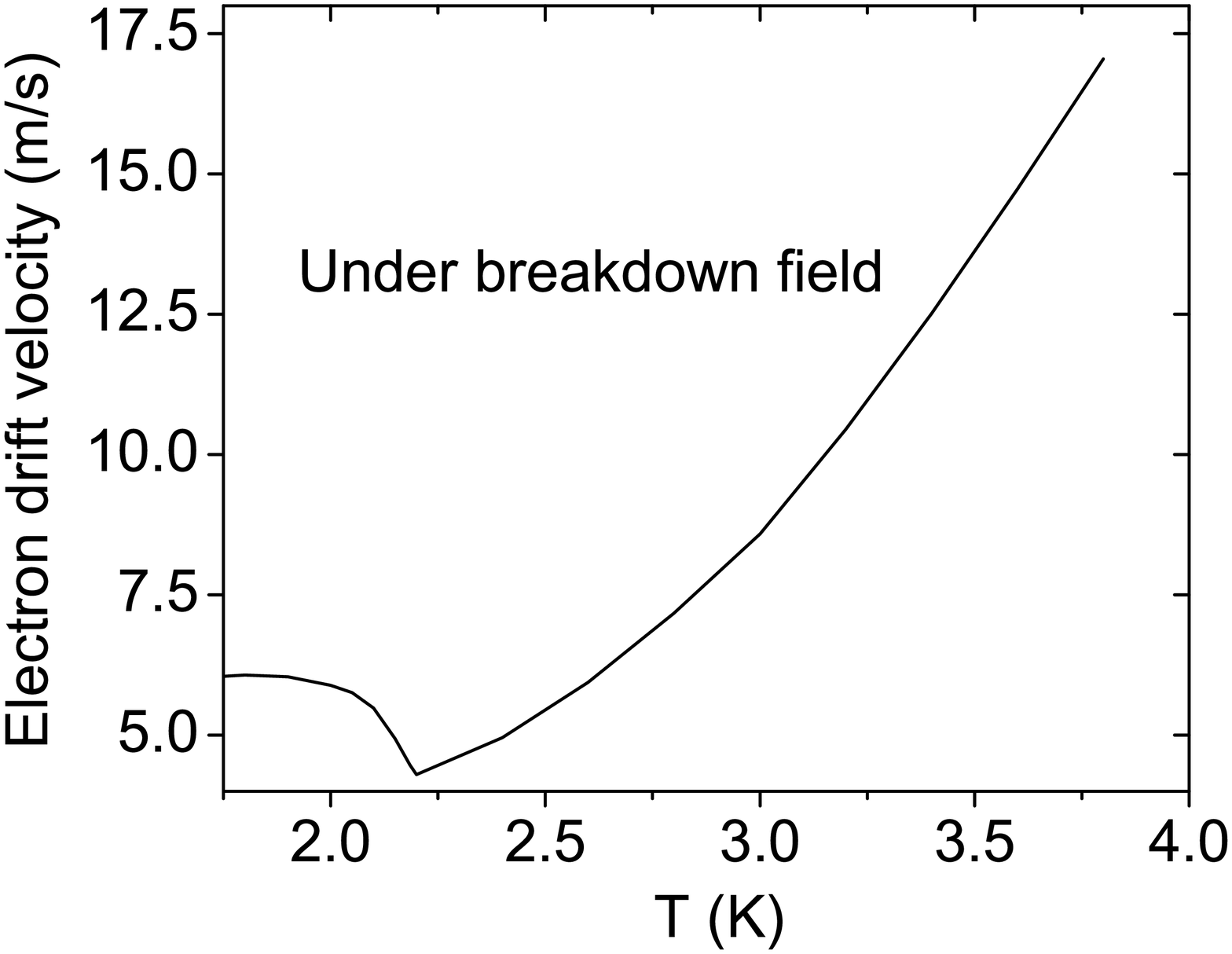}
\caption{(color online). The electron drift velocity in liquid helium as a function of the temperature under a drift field that is close to the breakdown field for saturated helium vapor.} \label{Electron-speed}
\end{figure}

The drift speed of the electrons $v$ in liquid helium is given by $v={\mu}E_{\texttt{drift}}$, where $\mu$ is the electron mobility in the liquid. Electrons drift much slower in liquid helium than in liquid xenon and liquid argon due to the small $\mu$ associated with the bubble state in helium. Based on the known mobility of the electrons in helium~\cite{Donnelly-mobility}, their drift velocity under a field close to the vapor breakdown field as given in Fig.~\ref{Breakdown field}~(b) can be calculated. The result is shown in Fig.~\ref{Electron-speed}. The electron speed drops with increasing temperature below the lambda point ($\sim2.17$~K), which is due to the steep drop of the electron mobility. Above the lambda point, $\mu$ decreases slowly with increasing temperature, and the electron speed rises steadily as the breakdown field increases. At 3.2~K, the electron speed is about 10~m/s. For a 10-liter sensitive volume, if we take the distance from the bottom electrode to the liquid/gas interface to be 20~cm, the maximum delay time between S1 and S2 will be about 20 ms. For a well shielded detector, event pileup should be well below the level that would cause mismatching of S1 and S2 signals.

Alternatively one may detect the extracted electrons using Gas
Electron Multipliers (GEMs) or Thick Gas Electron Multipliers
(THGEMs)~\cite{Buzulutskov-2005,Buzulutskov-2012, Breskin}.
Detecting electrons using GEMs has already been studied
experimentally by the ``e-bubble''
collaboration~\cite{Ju-2007,Ju-2007-Chin}. A disadvantage of the GEM
is that it gives poorer energy resolution than proportional
scintillation since it relies on a breakdown for electron gain.
However, because of the slow electron drift speed, the electrons
will arrive at the GEM one at a time, easily distinguishable due to
the excellent GEM timing resolution. We may operate the GEM in a
single electron detection mode, counting single electron pulses
instead of using the pulse sizes to determine the event energy. Note
that it was shown that in ultrapure helium gas GEM can operate only
at charge gain close to unity~\cite{Galea,Miyamoto}. However, during
the avalanche development, excited helium atoms and molecules are
produced. The decay of these electronic excitations leads to the
emission of 16~eV photons. Instead of detecting the charge produced
in the GEM, we may detect these photons with arrays of PMTs. At the
same time the GEM (or stack of GEMs) could also be used to amplify
the prompt scintillation signal. The side of the GEM facing the
liquid could be coated with Cesium Iodide (CsI)~\cite{Breshin-CsI}
or other photocathode material so as to be sensitive to the prompt
scintillation light. Furthermore, the extracted electrons may also be detected via electroluminescence
produced under very high ($\sim$1-10~MV/cm) fields on the surface of
thin wires or a GEM immersed in liquid helium. Such
electroluminescence has already been observed in liquid
Ar~\cite{Lightfoot} and liquid Xe~\cite{XENON100}. This could allow
electrons to be individually detected, while not subjecting gaseous
helium to a strong electric field.

\subsection{Low temperature operation scheme}
At very low temperatures (e.g. 100~mK), calorimetric sensors with
small heat capacity can be used for particle/photon detection with
remarkable sensitivity and low threshold. A significant advantage of using
calorimetric sensors is that one could in principle cover all the
inner surface of the detector with calorimetric sensor arrays, while not being limited by the 20--30\% quantum efficiency typical of photocathodes. S1
light collection efficiency approaching 100\% might be achieved, which
would improve the rejection power of the detector. At low helium temperatures, the thermal coupling between the calorimetric sensors and the liquid helium is weak, enabling them to be immersed in the liquid helium without losing much thermal signal to the bath.

The field of low temperature bolometers is developing rapidly. The types of
temperature sensors most commonly used are neutron-transmutation
doped (NTD) Germanium thermistors, and superconducting transition
edge sensors (TES)~\cite{Alessandrello-1999, Irwin-2000}. The
electric conductivity of NTD sensors strongly depends upon the
temperature, with typically resistance of $1\sim100$~M$\Omega$ at
low temperatures. NTD thermistors are easy to use because they can
be operated with conventional electronics. Mass production is also
possible for the NTD sensors. A TES is a superconducting strip
operating at the temperature of its superconducting-normal
transition. The resistance in the normal state is usually
10~m$\Omega\sim1\Omega$, and the temperature dependence of the
resistance can be very large at the transition. Recent developments
include not only improvements in single TES's, but also large arrays of
TES's and techniques for multiplexing them. For this dark matter application, especially promising are non-equilibrium detectors, in which interactions produce quasiparticles in a superconducting strip, which can be then collected in a TES and detected \cite{Cabrera-2008}. Another possible
choice is metallic magnetic calorimeters
(MMC)~\cite{Bandler-1993, Kim-2004,Kim-thesis}. MMCs are based upon the use of
magnetic sensors to measure very small temperature changes resulting
from the absorption of energy by energetic photons. Instead of
measuring the resistance of a sensor such as an NTD or a TES, an MMC
measures the change of magnetization of paramagnetic ions in a
metallic host~\cite{Booth-1996, Pretzl-2000}. The internal thermalization
time of MMCs is very fast, within a microsecond.

\begin{figure}[htb]
\includegraphics[scale=0.55]{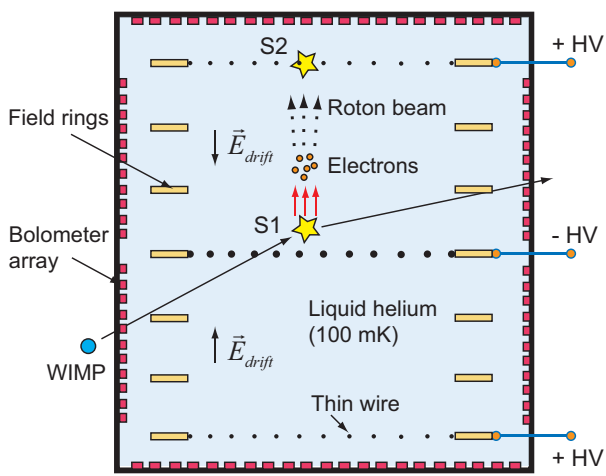}
\caption{(color online). A schematic of a single-phase helium detector operated at mK temperatures.} \label{mK-setup}
\end{figure}
A conceptual schematic of a calorimeter-based single-phase helium
time projection chamber is shown in Fig.~\ref{mK-setup}. The prompt
16~eV photons produced by a recoil event hit the bolometer arrays
and deposit their energy in the sensors which give rise to the S1 signal.
At low temperatures, bolometers may be made sensitive enough to
allow the detection of individual photons~\cite{Cabrera-2008, Kim-thesis}. To
extract the ionized electrons, three electrodes are arranged in a
way to drive the electrons toward either the top or the bottom
electrode. Since all the electrodes are immersed in liquid helium,
high voltages ($\sim\pm100$~kV) can be applied to them. The drift
field with the shown electrode arrangement can be made as high as a
few tens of kV/cm, which helps to improve the rejection power of the
detector. To amplify and detect the charge signal, both the top and
the bottom electrodes are made of thin wire arrays. The extracted
electrons drifting towards these thin wires may produce
electroluminescence as they approach the wire surface. These photons
(S2) can be detected by the same bolometer arrays. Again, event
position reconstruction can be made based on the delay between S1
and S2, and the hit pattern on the top (or bottom) bolometer array.

Note that the mobility of electron bubbles in liquid helium
increases drastically with decreasing temperature. Under saturated
vapor pressure, if the velocity of the electrons exceeds a threshold
velocity of the order of $40$ m/s, quantized vortex rings are
nucleated. The electrons can get stuck on them, and the charged
vortex ring moves through the liquid as a single entity. A striking
feature of the electron-ring complex is that its velocity decreases
with increasing energy~\cite{Donnelly}. When a strong drift field (a
few kV/cm) is applied, the velocity of the charged vortex rings can
be as low as on the order of 1~m/s~\cite{Bruschi-1966}. However, it
has been shown that at low temperatures when isotropically purified
helium is pressurized to above 15 bar, electrons can be driven at a
speed close to or higher than the Landau velocity ($\sim50$~m/s).
Instead of nucleating vortex rings, the electrons spontaneously emit
roton pairs~\cite{Phillips-1974, Allum-1976, Nancolas-1985}. The
rate of roton emission depends on the field strength. Furthermore,
it has been shown that when the electron speed is not too much
higher than the Landau velocity, the majority of the emitted rotons
tend to have momentum aligned in the same direction with the
electron velocity~\cite{Bowley-1985}. A roton beam is formed
accompanying every extracted electrons. Note that rotons in the
R$^+$ branch move along the electron velocity direction due to
their positive effective mass, while R$^-$ rotons are emitted in the
opposite direction since their effective mass is negative in
helium~\cite{Tucker-1985}. Despite the low transmission coefficient
of the roton energy across the liquid-solid interface at low temperatures, a
fraction of the roton energy can nevertheless transmit into the bolometers
and be detected~\cite{Tanatarov-2010}. Detecting the rotons provides
a unique way for charge signal amplification and detection, with potentially hundreds of keV of roton signal produced by each drifted electron. Operation at pressure \(>15\)~bar, as required in this approach, may also be advantageous for maintaining higher drift field by suppressing gas bubble formation.

Note that in the initial proposal by Lanou \emph{et al.}~\cite{Lanou-1988}, the idea of charge signal amplification via roton/vortex generation was briefly mentioned. It was proposed that event discrimination might be achieved based on the ratio of prompt rotons accompanying the initial recoil deposition to the delayed rotons from the drifted charge. Due to the low transmission of roton energy into the bolometer surface, detecting the prompt rotons for low energy recoils can be quite challenging. Nevertheless, if detection of prompt rotons and phonons could be accomplished for very low energy nuclear recoil events, search for extremely light WIMPs may be conducted. At very low energies, ionization is strongly suppressed for nuclear recoil events, and even a single electron from an electron recoil event would produce a large roton signal, allowing electron recoil backgrounds to be vetoed. In addition, the roton/phonon ratio may be different for nuclear and electron recoils, allowing electron recoil backgrounds to be discriminated by bolometer pulse shape.

It is worthwhile mentioning that at low temperatures, metastable
helium molecules in triplet state can drift at a speed of a few
meters per second~\cite{Zmeev-2012}. When these molecules collide on
the bolometer surface, they undergo non-radiative quenching and
release over 10~eV of energy depending on the material of the impinged
surface. A significant amount of this energy will be deposited into
the calorimetric sensor, which may allow us to detect the molecule
signal (S3). A combined analysis of S2/S1 and S3/S1 ratios may
further improve the rejection power of the detector.

\section{Sensitivity}
In any direct dark matter detection scheme, the primary requirement
is the strong reduction of radioactive backgrounds. The approaches
described above are designed to enable this, since liquid helium may
readily be purified of impurities, and because the ratios of signal
channels may be used to identify gammas and betas on an
event-by-event basis.

Internal backgrounds (due to radioactive impurities within the
target material) are particularly straightforward to eliminate in
liquid helium. Like other noble gases, helium is readily purified
with heated getters to remove any chemically reactive species, which
includes anything that is not a noble gas. In addition, helium has
no long-lived radioactive isotopes and therefore has no intrinsic
backgrounds, unlike argon and krypton which contain the beta
emitters $\rm ^{39}Ar$ and $\rm ^{85}Kr$. Activated charcoal
adsorber may then be used to remove all other radioactive noble
gases (e.g. radioactive argon, krypton, and radon), since all other
noble gases have larger polarizabilities and masses, leading to
larger binding energies to charcoal and substantially larger
adsorption coefficients \cite{McKinsey-2000, Harrison-2006}. Getters
and charcoal may be used to purify the helium after it is
transported underground. Any remaining impurities will fall out of
liquid helium and stick to plumbing and detector walls, and any
beta, alpha, or nuclear recoil background due to these impurities
may be removed through position reconstruction.

External backgrounds may be reduced through shielding, careful
detector materials selection, and event discrimination. The dominant
background in this experiment is expected to arise from gamma rays
that Compton scatter in the helium and produce low-energy events
that could be confused with dark matter particles. This gamma ray
Compton scattering background tends to be flat with energy at low
energies, and a typical Compton scattering background rate for a
shielded dark matter experiment is about 1 event/keVee/kg/day,
though this may be reduced further with special care given to
shielding and detector materials. Neutron backgrounds are typically
well subdominant to gamma rays, for most detector construction
materials. Radon daughter backgrounds on inner detector surfaces can
be troublesome, but are eliminated in this scheme through the
excellent position reconstruction inherent to noble liquid time
projection chambers.

With gamma rays generating the dominant background, it is crucial to
have excellent electron recoil versus nuclear recoil discrimination.
From the quantitative estimates described in Section II above, we
have good confidence that liquid helium will indeed provide
excellent discrimination power. In addition, the helium detector may
surrounded with a veto detector that is sensitive to gamma rays.
Using organic scintillator, liquid xenon, or liquid argon as
detector materials, such a veto may be used to efficiently tag gamma
rays that small-angle-scatter in the helium fiducial volume, produce
low energy events, and escape. Efficiencies of 90-98\% may be
expected, as predicted for the DarkSide and LUX-ZEPLIN experiments
\cite{Darkside, LZ}.

\begin{figure}[htb]
\includegraphics[scale=0.45]{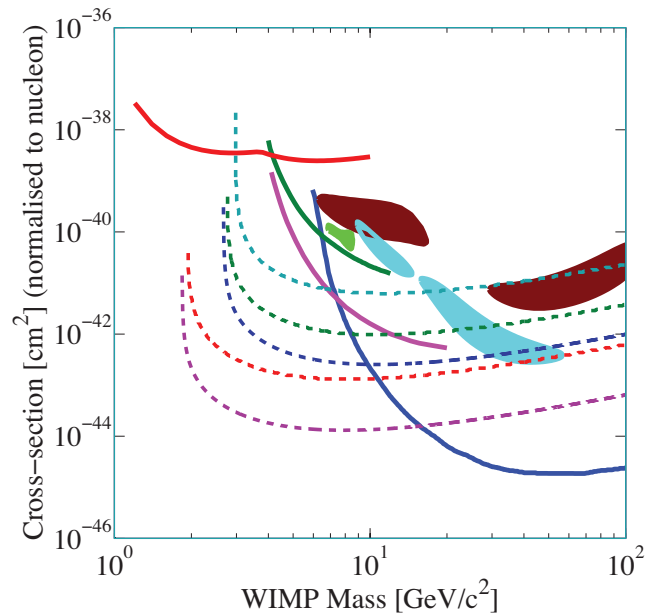}
\caption{(color online). Spin-independent WIMP exclusion curves (solid lines), potential WIMP signals (solid regions), and projected liquid helium 90\% sensitivity curves (dashed lines) in the region of 1-100 GeV WIMP mass. Exclusion curves include DAMIC in red \cite{DAMIC}, CDMS-II in green \cite{CDMS-lite}, XENON10 in magenta \cite{XENON10}, and XENON100 in blue \cite{XENON100}. Potential WIMP signals include DAMA in red \cite{DAMA}, CRESST in light blue \cite{CRESST}, and CoGeNT in green \cite{COGENT}. Projected liquid helium sensitivities for different detector parameters are shown as dashed lines, including light blue (10 kV/cm, 20\% S1 collection, 4.8 keV threshold), green (20 kV/cm, 20\% S1 collection, 4.4 keV threshold), blue (40 kV/cm, 20\% S1 collection, 4.2 keV threshold), red (20 kV/cm, 80\% S1 collection, 2.8 keV threshold), and magenta (40 kV/cm, 80\% S1 collection, 2.6 keV threshold). Predicted limits assume an electron recoil background of 1 event/keVee/kg/day and a 95\% efficient gamma ray veto.} \label{Exclusion curves}
\end{figure}

As explained in section I, a useful figure of merit for light WIMP
searches is (nuclear mass)$\times$(energy threshold), which must be
minimized to get the best light WIMP sensitivity. In the case of
liquid helium, this must be balanced with the background reduction
achieved through discrimination of electron recoil events, which
improves with higher energy. Given helium's large predicted nuclear
recoil signals and  excellent discrimination, we expect an energy
threshold of about 4-5 keV with photomultiplier readout, potentially
reducible to 1-3 keV with bolometric readout. The low nuclear mass
of helium then gives access to very low WIMP masses, while still
having significant background reduction through discrimination.

While liquid helium will not provide significant self-shielding
against gamma rays and neutrons (a significant background rejection
method in LXe and LAr detectors), a plausible background rate of
$10^{-3}$ events/day/keVee/kg after discrimination will allow
excellent sensitivity to light WIMPs, for which current experimental
sensitivities are relatively weak. A detailed discussion of the background of a helium detector designed for the HERON project was given by Huang \emph{et al.}~\cite{Huang-2008}. For a 1 kg helium fiducial mass,
20\% light collection, a 20 kV/cm drift field, an energy threshold
of 4.8 keV, 300 days of operation, and a 95\% efficient gamma ray
veto, one background event is predicted, with a WIMP-nucleon
cross-section sensitivity of $10^{-42}$ cm$^{2}$ at 5 GeV, the dark
matter mass predicted by asymmetric dark matter models. Sensitivity
may be improved further with higher drift fields, more efficient
light collection, and larger exposure, potentially reaching $\rm
10^{-44}$ cm$^{2}$ or better between 2-20 GeV. Some predicted light
WIMP sensitivities are summarized above in Figure~\ref{Exclusion
curves}.

\section{Conclusion}
We conclude that liquid helium is an intriguing material for the direct detection of light WIMPs, as it combines multiple signal channels, comparatively large signals for nuclear recoils, a low target mass, and the capacity for electron recoil discrimination. In the detector schemes proposed here, a high electric field is used to extract electrons from nuclear recoil tracks, allowing a sizable charge signal, time projection chamber readout, and good position resolution. Before dark matter experiments can be performed with this technology, a method of detecting single electrons in liquid superfluid helium must be demonstrated. In addition, detailed measurements must be done of the nuclear and electron recoil signal and discrimination efficiency at low energies.

\begin{acknowledgements}
We thank R.E. Lanou, D. Hooper, D. Prober, T.M. Ito, G.M. Seidel, and A. Buzulutskov for valuable discussions and comments, and D. Klemme for her assistance in preparation of this manuscript.
\end{acknowledgements}

\end{document}